77 New Thermodynamic Identities among Crystalline Elastic Material Properties

Leading to a Shear Modulus Constitutive Law in Isotropic Solids

by

S. J. Burns

Materials Science Program

Department of Mechanical Engineering

University of Rochester

Rochester, NY 14627

May 28, 2018

Abstract

Thermodynamics of crystalline materials is analyzed using strain volumes, an incremental tensorial state variable which is the volume per unit mass multiplied by the incremental strain. It is shown that the derivatives of the strain volume variables with respect to the stresses reduces to conventional well known isothermal, elastic, crystallographic compliances and crystallographic thermal expansion coefficients. The formulism is extended to all third order derivatives which establish 77 new thermodynamic identities: 27 are thermal and 50 are among selected stress components of elastic compliances. The stress dependence of the heat capacity is typically coupled into the crystallographic thermal expansion coefficient's temperature dependence; the temperature dependence of the elastic compliances is related to the stress dependence of the thermal expansion coefficients; stress dependent relationships among elastic compliances are also found. The paper emphasizes internal self-consistency. General triclinic identity relations which are of academic interest are applied to specific higher symmetries including orthorhombic, tetragonal and cubic; identities in higher symmetry crystals are very useful. A new generalized, constitutive law is found and applied to isotropic polycrystalline materials; the concept is based on zero shear thermal expansion coefficients. The constitutive modulus relation describes thermal and pressure properties only through an isochoric modulus.

*Introduction*

Thermodynamics of solids is an old subject that has been given new life by recent experiments on solid materials at extreme stresses, temperatures and pressures [1, 2]. Fusion's energy goal is to compress deuterium to 1/20 of its volume so it can 'burn'. Thus, the experiments are performed at extreme pressures. The mechanical work term in the first law of thermodynamics is often small at atmospheric pressures when compared to the heat term for solids. Materials under stresses or pressures of 100 TPa will add significant mechanical contributions to the internal energy of solids. When deformation occurs at 100 TPa the mechanical work is nine orders of magnitude larger than at atmospheric pressure.

It has been recently found experimentally that at very high pressures such as obtained in laser driven shock compression, laser ramp loading or diamond anvil cells that aluminum has several high pressure phases [3-5] including HCP and BCC. Potassium and sodium have very special electronic phases [6-9] and double hexagonal hP4 structure where the electrons released from atoms by pressure are collected in the interstices of the crystalline structures. Silicon [10, 11] has many of the analytically predicted high pressure phases and Fe has structures [12, 13, 14] not known at atmospheric pressures and engineering stresses.

Determination of the internal energy of a solid forms the basis for prediction of equations of state for many solid materials. In molecular dynamics, density functional theory, particle in cell, and other electron-electron or electron-atomic interactions the forces and atomic motions are used to predict the internal energy. These calculations are generally for *ab initio* density functional methods and determine the total energy of a crystal using for example the Vienna Ab-initio Simulation Package, VASP or from first principles methods, such as density functional theory based quantum molecular dynamics, path-integral Monte Carlo and quantum Monte Carlo methods which have been developed over the past decades to understand the properties of materials under extreme conditions [15, 16, 17]. Such models are approximations to real crystals and predict thermodynamic properties that are often difficult to experimentally verify and measure. Typically, the number of atoms in *ab-initio* and first principle systems is generally fixed so the total energy in the system is energy per unit mass. The atom to atom interactions and constraints are always holonomic; atomic and electron magnetic moments and quantum effects are sometimes ignored. The temperature is often zero. The interaction potential is a total description from all atomic interactions but have no mechanical work because the boundaries are fixed.

Stress and strain variables are frequently used to find the energy per unit volume for materials [18-20]. *Nota bene* that references [18] and [19] have special caveats about reference states since energy per unit volume is used. The system mandates addition or removal of atoms from the thermodynamic system when there are dilatational strain tensor components. Atoms need to be added or removed from the system at the correct chemical potentials. The stress and strain state variables then require inclusion as chemical potentials for all dilatational strains at the appropriate conditions. This complicates rather than simplifies the system's energy. J. Willard Gibbs [21] recognized that energy per unit volume did not give a valid thermodynamic description of a stressed solid. The reason is that the volume per unit mass adds the additional state variable to the six independent stress variables plus the temperature [22, 23]. Examples where volume per unit mass is an additional thermodynamic variable are often from geological materials at pressures found within the Earth and Solar System. Examples include the total energy per unit volume converted to energy per unit mass by multiplying by the volume per unit mass see: $MgSiO_3$-perovskites [24], MgO [25], $SiO_2$ [26], and many other works.

Most scientists and engineers have excellent understanding and feeling for the energy of systems. The derivatives of the energy are the state variables in the system; the second derivatives of the energy are the physical properties; however, there is very little understanding and physical intuition of the third derivatives of the system's energy. Derivatives of properties and their interrelationships are the main subject of this paper. This paper emphasizes mostly self-consistent physical properties based on thermodynamic identities found from the third-order derivatives of a Gibbs like energy function. There are only predictions of physical properties aside from internal self-consistency, at the end of the Conclusions. A general description of strain volume thermodynamics [27-29] is introduced and applied to a crystal of triclinic symmetry. The higher order symmetries of orthorhombic, tetragonal and cubic are subsequently used as specific examples to find interrelationships among stress and temperature dependencies of elastic compliances. Tables of Jacobian algebra [30-31] are used throughout so the isothermal and iso-piezo properties described here can be easily changed to isochoric or adiabatic properties for other measured physical conditions. Examples of transformations are used throughout the text; the transformation mathematics is included in Appendix A. Finally, a 'constitutive equation' is found for isotropic polycrystalline solids based on the form of the third order Gibbs function shear compliance property identity.

*Strain Volume Thermodynamics*

The incremental first law of thermodynamics using energy per unit mass in a stressed solid system is

$$du = Tds + v\sigma_{ij}d\varepsilon_{ij} \qquad (i, j = 1, 2, 3) \qquad (1)$$

$u$ is the internal energy of the system per unit mass, $T$ is the absolute temperature, $s$ is the entropy per unit mass, $v$ is the volume per unit mass, $\sigma_{ij}$ is the tensorial stress and $\varepsilon_{ij}$ is the tensorial strain. The first incremental term on the right is the heat added to the system and the second term is the incremental work done on the stressed solid per unit mass. The total mechanical work done on the system is found by summing the tensors for all repeated indices from 1 through 3. The incremental mechanical work term in equation (1) is in units of energy per unit mass. Equation (1) is identical to the first law used in reference [23] and just about everywhere.

In what follows below, the strain volume is obtained by combining $v$ with the incremental strains. The incremental strains and the volume are thus 6 thermodynamic variables. In this way a new variable, the strain volume variable, $\Omega$, is defined with units of volume per unit mass. $\Omega$ when multiplied by the stress, $\sigma$, with units of force per unit area or energy per unit volume gives a product of the two terms with energy per unit mass.

$$d\Omega_{ij} \equiv v d\varepsilon_{ij} \qquad (2)$$

The strain volume is a tensor quantity which reduces to $dv$ when the strain field is pure dilatation. In this case, we see for a fluid like solid that

$$d\Omega_{ii} = v d\varepsilon_{ii} = v(d\varepsilon_{11} + d\varepsilon_{22} + d\varepsilon_{33}) = v(\frac{dv}{v}) = dv \qquad (3)$$

The dilatational components of the trace of the strain tensor are the incremental change of the volume over the volume as is well known. The hydrostatic stress is just the negative pressure so equation (1) with this restriction reduces to the fluid description of the first law of thermodynamics. The mechanical work will now be written using strain volume

$$v\sigma_{ij} d\varepsilon_{ij} = \sigma_{ij} d\Omega_{ij}. \qquad (4)$$

The tensorial stresses and strains are reduced to vectors and the elastic compliance tensors to matrices following the work by Voigt [22] and later adapted by Nye [19]. The strain matrix contains factors of 2 and 4 as is standard practice introduced by Voigt. See reference [19] for the reductions rules for strain contractions from tensors to vectors. Thus equation (1) becomes

$$du = Tds + \sigma_i d\Omega_i \qquad \text{($i$ goes from 1 to 6)} \qquad (5)$$

There are 7 independent variables in the above equation, entropy and 6 strain volumes. Equation (5) is transformed to a free energy with temperature and stress as the independent variables by:

$$g \equiv u - Ts - \sigma_i \Omega_i. \qquad (6)$$

The Gibbs like free energy in incremental form for a stressed solid is:

$$dg = -s dT - \Omega_i d\sigma_i \qquad (7)$$

The entries found in Table 1 are self-consistent descriptions of the physical properties using isothermal and isostress as the independent variables in equation (7). The following definitions are used in Table 1:

The heat capacity at constant stress,

$$C_\sigma \equiv T \left.\frac{\partial s}{\partial T}\right|_{\sigma_i}, \qquad (8)$$

the linear crystallographic thermal expansion coefficients,

$$\beta_i \equiv \frac{1}{v}\left.\frac{\partial \Omega_i}{\partial T}\right|_{\sigma_i}, \qquad (9)$$

and the isothermal elastic crystallographic compliance,

$$S_{ij} \equiv \frac{1}{v}\left.\frac{\partial \Omega_i}{\partial \sigma_j}\right|_{T,\sigma_{j'}}. \qquad (10)$$

Table 1 also contains the Maxwell relations: this makes Table 1 symmetric about the property diagonal because the second derivatives of $g$ taken in either order are equal i.e., Maxwell's relations. Table 1 is very similar to the elastic descriptions for crystals that have been well known since Voigt first published his book on crystal elasticity [22]. It is seen that $v$ is displayed in every entry except for the heat capacity entry in the upper left. The reason is that the incremental strain volume variable contains $v$. Table 1 thus contains $v$ in all entries aside from the heat capacity which is already in units of mass; the incremental strain in the strain volume definition keeps the well-known symmetry of the thermal expansion and elastic compliance found when stress, strain and temperature are the state variables.

The third order derivatives of $g$ are listed in Table 2; they relate properties of the heat capacity, thermal expansion coefficients and elastic compliances. The third order derivatives are of the free energy and are quite different than just additional derivatives of the elastic compliances which usually are called 'third order elastic compliances' which do not include any thermal terms. Third order elastic compliances typically describe [32] the non-linear elastic properties of crystals. Table 2 however is very unique when compared to what is in the thermodynamic literature for elastic crystals. This is because it describes the derivatives of the Gibbs like free energy and not just the higher order elastic compliances. The general case is for a triclinic crystal; Table 2 uses a notation closely related to the third order elastic compliances [33] with one major exception: the temperature derivatives are assigned the symbol, '0'. Again, like the Maxwell relations in Table 1 derivatives taken in any order are equal. The independent variables in Table 2 are still temperature and the six stresses. All the elastic terms in Table 2 come originally from six rank tensors. The tensor reductions to matrix format are included in Table 2. Within Table 2 the Maxwell type relations can be seen as symmetry about minor diagonals. There are three general types: those that contain two temperature terms and one stress, those with only 1 temperature term and two stresses, and finally terms with three stresses. Specific examples are seen in equations (11), (12) and (13).

$$-\frac{\partial^3 g}{\partial T \partial \sigma_1^2} = \frac{\partial^2 \Omega_1}{\partial T \partial \sigma_1} = \frac{\partial(vS_{11})}{\partial T}\bigg|_{\sigma_i} = \frac{\partial(v\beta_1)}{\partial \sigma_1}\bigg|_{T,\sigma_1'} = \frac{\partial^2 \Omega_1}{\partial \sigma_1 \partial T} = -\frac{\partial^3 g}{\partial \sigma_1 \partial T \partial \sigma_1}, \qquad (11)$$

$$-\frac{\partial^3 g}{\partial T \partial \sigma_1 \partial \sigma_2} = \frac{\partial^2 \Omega_2}{\partial T \partial \sigma_1} = \frac{\partial(vS_{12})}{\partial T}\bigg|_{\sigma_i} = \frac{\partial(v\beta_2)}{\partial \sigma_1}\bigg|_{T,\sigma_1'} = \frac{\partial^2 \Omega_2}{\partial \sigma_1 \partial T} = -\frac{\partial^3 g}{\partial \sigma_1 \partial T \partial \sigma_2}, \qquad (12)$$

and

$$-\frac{\partial^3 g}{\partial \sigma_1 \partial \sigma_2 \partial \sigma_1} = \frac{\partial^2 \Omega_1}{\partial \sigma_1 \partial \sigma_2} = \frac{\partial(vS_{12})}{\partial \sigma_1}\bigg|_{T,\sigma_1'} = \frac{\partial(vS_{11})}{\partial \sigma_2}\bigg|_{T,\sigma_2'} = \frac{\partial^2 \Omega_1}{\partial \sigma_2 \partial \sigma_1} = -\frac{\partial^3 g}{\partial \sigma_2 \partial \sigma_1^2} \qquad (13)$$

Table 2 is not a description of third order, non-linear elastic compliances [33] nor is it the third order expansion developed about zero stresses. Table 2 should be thought of as the coupling between the thermal and stress variables required by *g* being a state function in equation 7. In the general form, these relations are not obvious. In what follows, below applications to crystals of higher symmetry are taken by restricting many entries in the general table to be zero.

Equation (1) used here has *v* proceeding the strain energy term but *v* is a dependent variable contained in $\Omega$. The 2$^{nd}$, 3$^{rd}$ or higher order derivatives must contain terms that involve the variable *v*. The stress derivatives of $\Omega$ don't directly involve *v*. In Table 2 given below the *v* terms are embedded in the $c_{ijk}$ terms presented here. Table 2 uses the notation related to 3$^{rd}$ order elastic compliances [32] but the entries are third order derivatives of the Gibbs like free energy with respect to the stresses and temperature variables.

Table 1 and 2 are both in Jacobian format. This permits all derivatives of state variables to be found containing only entries from the tables. Appendix A gives an example where the use of MatLab has been employed to evaluate the large matrices found here. Consider for example evaluation of the difference between the isostress heat capacity defined in Table 1 less the isometric heat capacity. This is written as:

$$c_\sigma - c_\Omega = (c_\sigma - T\frac{\partial s}{\partial T}\bigg|_\Omega) = (c_\sigma - T\frac{J(s,\Omega_1,\Omega_2,\Omega_3,\Omega_4,\Omega_5,\Omega_6)}{J(T,\Omega_1,\Omega_2,\Omega_3,\Omega_4,\Omega_5,\Omega_6)}) \qquad (14)$$

*J(s, $\Omega_1$, $\Omega_2$,.....,$\Omega_6$)* is the Jacobian entries read from the Table 1. The general case with all 21 elastic constants and 6 thermal expansion coefficients is over 20 pages of MatLab algebraic expressions. Using Appendix A and the table for cubic symmetry to evaluate the 7X7 Jacobian matrices ratio gives a simple result. N.B. the ratio of the $S_{ij}$'s shown below is not simply the reciprocal of the compressibility:

$$c_\sigma - c_\Omega = 3Tv\beta_1^2 \frac{(S_{11} - S_{12})}{(S_{11}^2 - 2S_{12}^2)} \tag{15}$$

The temperature, volume and the thermal expansion coefficient squared are as expected. The elastic compliances shown are from Jacobian algebra.

A second example is to find the ratio of the heat to the mechanical work, isothermally at constant stress. In what is below, the mechanical work is only from the stress $\sigma_1$ and the system is not constrained to fixed boundaries. If this system were an ideal gas the magnitude of the heat to work ratio would be exactly 1. This ratio is

$$\frac{\text{incremental heat}}{\text{incremental work}} = \frac{T}{\sigma_1} \frac{\partial s}{\partial \Omega_1}\bigg|_{T,\sigma_1'} = \frac{T}{\sigma_1} \frac{J(s,T,\sigma_2,\sigma_3,\sigma_4,\sigma_5,\sigma_6)}{J(\Omega_1,T,\sigma_2,\sigma_3,\sigma_4,\sigma_5,\sigma_6)} \tag{16}$$

Again using Appendix A it is found for the triclinic crystal structure that

$$\frac{T}{\sigma_1} \frac{\partial s}{\partial \Omega_1}\bigg|_{T,\sigma_1'} = \frac{T\beta_1}{\sigma_1 S_{11}} \tag{17}$$

Equation (17) shows that at very high stresses the ratio becomes small while at very high temperatures the ratio is larger. $\beta_1$ and $S_{11}$ will in general both depend on all the independent variables in the system. Also equation (17) shows that the heat capacity is not contained in this ratio since the thermodynamic process is isothermal. Consider the material chosen to be copper at atmospheric stress and T = 1000 K. Then this ratio is 21,000 but at $\sigma_1$ = 100 TPa and T = 10,000 K using atmospheric and room temperature values for $\beta_1$ and $S_{11}$ the estimate would be 0.002. The conclusion is that very high stresses reduces the effects of heat in the thermodynamic system and that copper when solid is very poorly represented by any ideal gas concepts.

Orthorhombic Symmetry

Orthorhombic symmetry applied to the general triclinic crystal in Tables 1 and 2 further restricts the physical properties and the third order Gibbs derivatives. Table 3 gives the example for orthorhombic symmetry. The cell remains orthorhombic with changes in temperature so $\beta_4$, $\beta_5$ and $\beta_6$ are all zero and the elastic compliances are restricted to orthorhombic symmetry. The immediate implication is that the heat capacity at constant stress is independent of all the shear stresses. In addition $v\beta_1$, $v\beta_2$ and $v\beta_3$ are independent of the shear stresses. The crystal symmetry restricts the elastic compliances so there are 9 independent elastic compliances as shown in Table 3. This implies that in Table 2, adapted to orthorhombic symmetry, the entries for the additional 12 elastic compliances are all zero. Thus in each row where there is a zero elastic compliance the $c_{ijk}$ are zero. In Table 3, these zeros have not been included but the thermodynamic identities are as shown. Table 3 also shows that $vS_{44}$, $vS_{55}$ and $vS_{66}$ are only dependent on the shear stresses $\sigma_4$, $\sigma_5$ and $\sigma_6$ respectively. All the remaining elastic compliances are independent of shear stress. Finally, it should be noted that the temperature dependence of $vS_{44}$, $vS_{55}$ and $vS_{66}$ are all zero through third order coupling because $\beta_4$, $\beta_5$ and $\beta_6$ are all zero. These last expressions are commented on in the Conclusion section of this paper. For all the crystal structures discussed here and below there is a caveat: a stress application does not change the crystal structure. For example, the application of any shear stress formally changes an orthorhombic crystal into a monoclinic lattice structure.

Tetragonal Symmetry

Tetragonal symmetry is quite similar to orthorhombic but there are even fewer elastic compliances and thermal expansion coefficients. In the case of tetragonal symmetry there are two crystal classes: the symmetry groups are 4$mm$, $\bar{4}$2$m$, 422 and 4/$mmm$ was chosen for illustration. This structure has 6 elastic compliances in a tetragonal body centered cell. Table 4 shows the restrictions on the physical compliances and their third order derivatives. The zeros in the table are from the thermal expansion coefficients and zero elastic compliances when applied through Table 2. There is one additional zero that is seen in Table 4: it comes because $S_{44} = S_{55}$. The derivatives of $S_{44}$ as taken with respect to $\sigma_5$ is zero and $S_{55}$ as taken with respect to $\sigma_4$ is zero but since $S_{44} = S_{55}$ it follows that $S_{44}$ has no shear stress dependence as shown in the table. Furthermore, all the temperature dependence of the shear compliances is only through $v$. This comes about because the assumption is that the cell remains tetragonal with changes in temperature. All shear thermal expansion coefficients are necessarily zero.

Cubic Symmetry

Cubic symmetry applied to the bottom of Table 1 and Table 2 is given in Table 5. The application of stress would again alter cubic symmetry except for the application of hydrostatic pressure. The application of heat is assumed to keep a cubic structure as cubic. Furthermore, the application of the stresses $\sigma_1$, $\sigma_2$ or $\sigma_3$ would generally change cubic symmetry to tetragonal or orthorhombic symmetry or even reduce the symmetry further. However, in what follows below the crystal symmetry is assumed to remain cubic with the application of stresses. This implies that the thermal expansion coefficients $\beta_4$, $\beta_5$ and $\beta_6$ are not only equal but are also zero. The thermal expansion coefficients $\beta_4$, $\beta_5$ and $\beta_6$ in cubic symmetry are all taken to be zero as the crystal remains cubic with increasing temperature. This restricts the third order derivatives of the elastic compliances with respect to their temperature dependence. The restrictions in Table 5 are obtained by setting non-cubic elastic compliances to zero and zero shear thermal expansion coefficients. The third order identity derivatives have significant consequences. Table 5 is the results of setting these zeros in the cubic description into the general $c_{ijk}$ entries from the Table 2.

*Conclusions*

Single crystal strain and temperature properties have been described through a third order Gibbs function using strain volumes. The third order derivatives in Table 2 have embedded in them 77 new thermodynamic identities of which 27 are thermal and 50 are among selected stress components on elastic compliances. This implies that when selecting physical properties to model or describe a crystal that there are many complex interactions. The Gibbs function third order derivatives are not easy to choose as shown in Tables 3-5. The extensive numbers of zero entries under shear interactions are restrictions on physical properties. Most restrictions are a direct result of the zero shear thermal expansion coefficients. Even non-linear shear properties of the shear compliances for cubic and tetragonal crystals are not allowed. The normal stresses are less restricted with non-linear terms possible.

The physical modeling of atomic interactions widely uses non-linear interactions in potentials and density functional theories. In the crystal classes investigated the temperature and stress dependence of the compliances are often reported to be the same as seen in Tables 4 and 5 with $S_{11}$, $S_{12}$ and $S_{13}$ having the same temperature and stress dependences. Here restrictions are a result of the third order Gibbs like energy function and not a consequence of DFT calculations or modeling.

One result that is a surprise is that for orthorhombic, tetragonal and cubic materials that the temperature dependence of the shear compliances is determined by the temperature dependence of the crystal volume alone. The experimental values for elastic shear compliances are most often from wave propagation; the wave propagates quickly so the condition is adiabatic and most experimental values are from adiabatic measurements. For the shear compliance say for $S_{44}$, for example we have equal adiabatic and isothermal compliances.

$$\frac{1}{v}\frac{\partial \Omega_4}{\partial \sigma_4}\bigg|_{T,\sigma_4'} = \frac{1}{v}\frac{\partial \Omega_4}{\partial \sigma_4}\bigg|_{s,\sigma_4'} = S_{44}. \tag{18}$$

In equation (18) use has been made of the shear thermal expansion coefficients being zero; the crystal cells remain orthogonal, tetragonal or cubic with changes in temperature. See [37] for a detailed discussion of the basis of equation (18).

The energy in a crystal at zero stress is 2/3 weighted to the shear phonons. In cubic, tetragonal and orthorhombic crystal structures it is clear that $\beta_4$, $\beta_5$ and $\beta_6$ are all zero. The Maxwell's relation entries for $s$ in Tables 3-5 show that it is independent of all shear stresses. This implies that the shear phonons can't couple into the shear elastic compliances. The concept of elastic shear phonons representing the entropy and thus a major part of the heat capacity of a solid is a foundation that is irrefutable. Yet, with the shear thermal expansion coefficients being zero, thermodynamics must restrict the temperature dependence of the shear elastic compliances as noted above. In Tables 3 through 5 when the shear strains $\Omega_4$, $\Omega_5$ and $\Omega_6$ are not zero, it is shown that $\beta_4$, $\beta_5$ and $\beta_6$ are still necessarily zero. The shear strains will change the crystal structure away from orthorhombic, tetragonal or cubic towards monoclinic or rhombohedral lattices; application of shear stress for monoclinic and rhombohedral cells would allow temperature and stress dependence in the shear compliances.

The implication of the identity in the last row of Table 5 is that

$$vS_{44} = \text{constant}_1 \tag{19}$$

In equation (19) $v$ is the volume and $S_{44} = 1/\mu_T$ is the engineering shear modulus (Lame's second constant). Equation (19) is not supported by experimental data but this conflict is resolved below and in another publication [37]. The equation developed below leads to the major contribution from this paper.

The expression that $vS_{44}$ in cubic crystals is independent of T and all stresses is not in agreement with experimental data. This apparent violation of a thermodynamic identity has led to a broader understanding of the role of shear stress and shear strains in describing material properties and the subsequent development of a constitutive law for the shear modulus. This is discussed below.

Equation (1) is an energy balance per unit mass; there is only direct mathematics and definitions between equation (1) and Tables 1-5 with the symmetry restrictions noted in Tables 2-5. If equation (1) were energy per unit volume rather than per unit mass as is used here, then again with no shear thermal expansion coefficients in structures of orthorhombic, tetragonal or cubic crystals we would then have $S_{44}$ being independent of $T$ and stresses. This is also not supported by experimental evidence.

Isotropic materials have only a single thermal expansion coefficient, so the $T$ dependence of $S_{11}$ and $S_{12}$ would necessarily be the same. Isotropic materials are not a crystal class but come about because of macroscopic averaging of microscopic polycrystalline properties within the material. Each crystallite in polycrystalline materials would show the thermal and elastic properties of its crystal symmetry class not those of isotropic materials. In polycrystalline materials isotropic properties prevail while each crystallite within the solid will have thermal expansion and contractions with the application of heat and stress. A shear stress applied to a polycrystalline solid will cause some crystallites to have normal stresses in some grains through grain boundary forces. These crystallites can thus change the intragranular temperatures with the application of shear stresses. In a single crystal of cubic material the application of $\sigma_4$, $\sigma_5$ or $\sigma_6$ is thought to not allow changes in temperatures. In the case where the material is polycrystalline then the material properties are not homogeneous and the thermodynamic descriptions will be phenomenological or constitutive. For example, it is easy to expect that the material would display zero shear thermal expansion coefficients only at zero shear stress while shear compliances are $T$ dependent.

It is instructive to restrict cubic materials even further so they can be considered 'isotropic.' Applied shear stresses, have a constitutive relation which describes non-shear thermal expansion coefficients since the cell is formally not a cubic structure with applied stresses. Yet phenomenologically it is easy to envision that $\beta_4$ is proportional to $\sigma_4$ and zero only when $\sigma_4 = 0$. The symmetries considered in the literature unfortunately don't allow that possibility.

Questioning the hypothesis of zero shear thermal expansion coefficients gives new insight into high symmetry behavior. In figure 1, the two isothermal shear lines shown cross at the point $\sigma_4 = 0$ and $\varepsilon_4 = 0$. The shear strains and shear stresses represents shear in a polycrystalline, isotropic material so the shear thermal expansion coefficients are zero only at a point. The shear thermal expansion coefficients are positive in the first quadrant and negative in the third quadrant in figure 1. At the crossing point they are zero. Equation (18) shows that the isothermal and adiabatic compliances are identical at a given temperature only at zero shear stress as was already discussed. Figure 1 also shows that adiabatic and isothermal compliances are only at the zero crossing point. Thus, equation (18) is suspect except at a single point. If equation (18) were true in general the constant entropy and the isothermal lines share would the same slope at all points. Even though the isotherms cross and therefore touch, the adiabatic lines can't touch as this would be a clear violation of Carathéodory's principle and a violation of the second law of thermodynamics. The point $\sigma_4 = 0$ and $\varepsilon_4 = 0$ therefore must be repulsive for all the adiabatic lines and they must turn away from this crossing point. This is only true for the shears stresses and strains not the dilatational strains as seen in Tables 3-5 which have entries in the upper left portions of the Tables and don't have isotherms intersecting. The isothermal lines for dilatation don't obviously touch and repulsive points among the dilatational stresses and strains are to be discussed in a separate publication [37].

Critical points [34] and repulsive points [35] share some similar characteristics. However, they are very different as critical points have an intrinsic instability; the many points in chemical phase diagrams and the spinodal have wide influences. Ma's work [34] shows how critical points results in power exponents. The clear violation of the thermodynamic identities found at the repulsive point in this paper and seen in equation (19) is found to be true provided

$$(v(T,p))^m \mu_T = \text{constant}_2 \qquad (20)$$

There is some support in the literature [36] that the shear compliance only depends on volume. Thus, considering the generalization of equation (19) to be equation (20) is not unreasonable. It is the full subject of [37]. Equation (20) is a constitutive function that contains only a single state variable in the shear modulus. The second state variable at the repulsive point is entropy due to shear that is always zero. The repulsive point contains all the temperatures as seen in figure 1 and all pressures which are independent of shear variables. Equation (20) will now be shown to be applicable to materials and very useful. $v$ is still the volume per unit mass and $\mu_T$ is the engineering, isothermal shear modulus (not the reciprocal of the tensorial, isothermal compliance which differs by a factor of 2). It should be noted that equation (20) can't just be based on shear thermal expansion coefficients being zero as this is incorrect in the literature. So equation (18) as derived is only true at a single point in the phase space and not of general interest. The reference [37] shows the generality of equation (20).

Figure 2 shows experimental data for polycrystalline, oxygen free copper in support of expression (20). The data was taken from NIST reference curves and their included tables in [38]. It is seen that equation (20) for copper is well supported by experimental data. About 25 additional plots of experimental data from metals, ceramics and minerals are located at this URL: SJB Web page URL. Figure 2 is not unique in establishing the validly of equation (20); the experimental data investigated supports equation (20) in many cases better than figure 2. In the case of copper temperature is the parametric variable while measurements were on an isobar for both $v$ and $\mu_T$. Reverting to classical thermodynamic notation for isotropic, polycrystalline solids it is also found that equation (20) yields for temperature and pressure derivatives the following expressions:

$$\frac{\partial \ln(\mu_T)}{\partial T} = -m(3\beta) = -m\frac{\partial \ln(v)}{\partial T}, \qquad (21)$$

where $3\beta$ is the volumetric thermal expansion coefficient. Figure 2 supports equation (20) while (21) relates the $T$ dependence of $\mu_T$ and $v$ through $3\beta$.

The derivative of equation (20) with respect to the pressure, $p$, gives:

$$-m\kappa_T \mu_T + \frac{\partial \mu_T}{\partial p} = 0, \qquad (22)$$

with $\kappa_T$ the isothermal compressibility. $\kappa_T$ and $\mu_T$ are related by Hooke's law for isotropic elastic solids [18] so,

$$-m\frac{3(1-2\gamma)}{2(1+\gamma)} + \frac{\partial \mu_T}{\partial p} = 0. \qquad (23)$$

$\gamma$ is Poisson's ratio. Taking $m$ from figure 2 and Poisson's ratio for Cu as 0.34, it is seen that

$$-2.8 + \frac{\partial \mu_T}{\partial p} = 0 \qquad (24)$$

Experimental data [40] reported on copper lists $d\mu_s/dp$ as 2.35. Again adiabatic and isothermal shear moduli are taken to be the same for shear. See [37].

The isothermal Figure 1 will not support equation (18) or (19) as shear thermal expansion coefficients are only zero at a single point. No additional data is included here but it should be noted that other polycrystalline materials also show excellent linear data when plotted as in figure (2) whereas data for equation (23) is well known in the geophysics literature [40].

Most geophysics interest is in the temperature and pressure dependence of isotropic materials [40]. Again, we would have $v^m \mu_T$ independent of $T$. All the temperature dependence in $S_{11}$ and $S_{12}$ must come from $\mu_T$ and $\gamma$. N.B. that the $v$ term drops when it is on both sides of most equations. Models of solids at high pressures are for isotropic materials with both pressure and temperature dependences often described. Table 5 shows how the thermal expansion coefficient couples into the temperature dependence of the compliances for cubic materials. The implications for isotropic materials are that under pressure when we have $\sigma_1$, $\sigma_2$ and $\sigma_3$ all equal and equal to $-p$, with $p$ being the pressure that all the temperature dependence is in $S_{44}$. In this case, Poisson's ratio in $S_{12}$ is considered independent of $T$. As noted above understanding single crystalline thermodynamics does not necessarily translate into polycrystalline isotropic properties. Isotropic materials have only 2 elastic compliances so $S_{44}$ and $S_{12}$ are chosen as being independent with $S_{11}$ the as a dependent quantity. There is only a single thermal expansion coefficient. Slightly changing Table 5 from cubic materials to isotropic materials gives the following conclusions: the implication is that all the temperature dependence would be in the shear modulus, $\mu_T$ through $v$ while Poisson's ratio would be independent of $T$ and stress. The same would be true for the pressure dependence of Young's modulus. There is analytic support for the pressure dependence of $S_{11}$ and $S_{12}$ being equal [41].

Table 1. Definitions of a thermodynamic system with $T$ and $\sigma_i$ as the independent variables in Jacobian format.

| f | $\partial/\partial T\vert_\sigma$ | $\partial/\partial\sigma_1\vert_{T,\sigma_1'}$ | $\partial/\partial\sigma_2\vert_{T,\sigma_2'}$ | $\partial/\partial\sigma_3\vert_{T,\sigma_3'}$ | $\partial/\partial\sigma_4\vert_{T,\sigma_4'}$ | $\partial/\partial\sigma_5\vert_{T,\sigma_5'}$ | $\partial/\partial\sigma_6\vert_{T,\sigma_6'}$ |
|---|---|---|---|---|---|---|---|
| g | -s | $-\Omega_1$ | $-\Omega_2$ | $-\Omega_3$ | $-\Omega_4$ | $-\Omega_5$ | $-\Omega_6$ |
| T | 1 | 0 | 0 | 0 | 0 | 0 | 0 |
| $\sigma_1$ | 0 | 1 | 0 | 0 | 0 | 0 | 0 |
| $\sigma_2$ | 0 | 0 | 1 | 0 | 0 | 0 | 0 |
| $\sigma_3$ | 0 | 0 | 0 | 1 | 0 | 0 | 0 |
| $\sigma_4$ | 0 | 0 | 0 | 0 | 1 | 0 | 0 |
| $\sigma_5$ | 0 | 0 | 0 | 0 | 0 | 1 | 0 |
| $\sigma_6$ | 0 | 0 | 0 | 0 | 0 | 0 | 1 |
| s | $C_\sigma/T$ | $v\beta_1$ | $v\beta_2$ | $v\beta_3$ | $v\beta_4$ | $v\beta_5$ | $v\beta_6$ |
| $\Omega_1$ | $v\beta_1$ | $vS_{11}$ | $vS_{12}$ | $vS_{13}$ | $vS_{14}$ | $vS_{15}$ | $vS_{16}$ |
| $\Omega_2$ | $v\beta_2$ | $vS_{12}$ | $vS_{22}$ | $vS_{23}$ | $vS_{24}$ | $vS_{25}$ | $vS_{26}$ |
| $\Omega_3$ | $v\beta_3$ | $vS_{13}$ | $vS_{23}$ | $vS_{33}$ | $vS_{34}$ | $vS_{35}$ | $vS_{36}$ |
| $\Omega_4$ | $v\beta_4$ | $vS_{14}$ | $vS_{24}$ | $vS_{34}$ | $vS_{44}$ | $vS_{45}$ | $vS_{46}$ |
| $\Omega_5$ | $v\beta_5$ | $vS_{15}$ | $vS_{25}$ | $vS_{35}$ | $vS_{45}$ | $vS_{55}$ | $vS_{56}$ |
| $\Omega_6$ | $v\beta_6$ | $vS_{16}$ | $vS_{26}$ | $vS_{36}$ | $vS_{46}$ | $vS_{56}$ | $vS_{66}$ |

Table 1. The above table contains definitions of the constant stress heat capacity, $C_\sigma$, the linear thermal expansion coefficients, $\beta_i$, and the elastic compliances, $S_{ij}$, for a general stressed crystal. It can be seen that there is one heat capacity at constant stress, 6 independent thermal expansion coefficients and 21 independent elastic compliances for 28 independent terms. The matrix of the physical properties is symmetric about the main diagonal because the second derivatives of $g$ may be taken in either order.

Table 2. Third order derivatives of g written in a Jacobian format with T and the six stresses, $\sigma_i$, as independent variables.

| f | $\partial/\partial T\|_\sigma$ | $\partial/\partial \sigma_1\|_{T,\sigma_1'}$ | $\partial/\partial \sigma_2\|_{T,\sigma_2'}$ | $\partial/\partial \sigma_3\|_{T,\sigma_3'}$ | $\partial/\partial \sigma_4\|_{T,\sigma_4'}$ | $\partial/\partial \sigma_5\|_{T,\sigma_5'}$ | $\partial/\partial \sigma_6\|_{T,\sigma_6'}$ |
|---|---|---|---|---|---|---|---|
| $c_\sigma/T$ | $c_{000}$ | $c_{001}$ | $c_{002}$ | $c_{003}$ | $c_{004}$ | $c_{005}$ | $c_{006}$ |
| $v\beta_1$ | $c_{001}$ | $c_{011}$ | $c_{012}$ | $c_{013}$ | $c_{014}$ | $c_{015}$ | $c_{016}$ |
| $v\beta_2$ | $c_{002}$ | $c_{012}$ | $c_{022}$ | $c_{023}$ | $c_{024}$ | $c_{025}$ | $c_{026}$ |
| $v\beta_3$ | $c_{003}$ | $c_{013}$ | $c_{023}$ | $c_{033}$ | $c_{034}$ | $c_{035}$ | $c_{036}$ |
| $v\beta_4$ | $c_{004}$ | $c_{014}$ | $c_{024}$ | $c_{034}$ | $c_{044}$ | $c_{045}$ | $c_{046}$ |
| $v\beta_5$ | $c_{005}$ | $c_{015}$ | $c_{025}$ | $c_{035}$ | $c_{045}$ | $c_{055}$ | $c_{056}$ |
| $v\beta_6$ | $c_{006}$ | $c_{016}$ | $c_{026}$ | $c_{036}$ | $c_{046}$ | $c_{056}$ | $c_{066}$ |
| $vS_{11}$ | $c_{011}$ | $c_{111}$ | $c_{112}$ | $c_{113}$ | $c_{114}$ | $c_{115}$ | $c_{116}$ |
| $vS_{12}$ | $c_{012}$ | $c_{112}$ | $c_{122}$ | $c_{123}$ | $c_{124}$ | $c_{125}$ | $c_{126}$ |
| $vS_{13}$ | $c_{013}$ | $c_{113}$ | $c_{123}$ | $c_{133}$ | $c_{134}$ | $c_{135}$ | $c_{136}$ |
| $vS_{14}$ | $c_{014}$ | $c_{114}$ | $c_{124}$ | $c_{134}$ | $c_{144}$ | $c_{145}$ | $c_{146}$ |
| $vS_{15}$ | $c_{015}$ | $c_{115}$ | $c_{125}$ | $c_{135}$ | $c_{145}$ | $c_{155}$ | $c_{156}$ |
| $vS_{16}$ | $c_{016}$ | $c_{116}$ | $c_{126}$ | $c_{136}$ | $c_{146}$ | $c_{156}$ | $c_{166}$ |
| $vS_{22}$ | $c_{022}$ | $c_{122}$ | $c_{222}$ | $c_{223}$ | $c_{224}$ | $c_{225}$ | $c_{226}$ |
| $vS_{23}$ | $c_{023}$ | $c_{123}$ | $c_{223}$ | $c_{233}$ | $c_{234}$ | $c_{235}$ | $c_{236}$ |
| $vS_{24}$ | $c_{024}$ | $c_{124}$ | $c_{224}$ | $c_{234}$ | $c_{244}$ | $c_{245}$ | $c_{246}$ |
| $vS_{25}$ | $c_{025}$ | $c_{125}$ | $c_{225}$ | $c_{235}$ | $c_{245}$ | $c_{255}$ | $c_{256}$ |
| $vS_{26}$ | $c_{026}$ | $c_{126}$ | $c_{226}$ | $c_{236}$ | $c_{246}$ | $c_{256}$ | $c_{266}$ |
| $vS_{33}$ | $c_{033}$ | $c_{133}$ | $c_{233}$ | $c_{333}$ | $c_{334}$ | $c_{335}$ | $c_{336}$ |
| $vS_{34}$ | $c_{034}$ | $c_{134}$ | $c_{234}$ | $c_{334}$ | $c_{344}$ | $c_{345}$ | $c_{346}$ |
| $vS_{35}$ | $c_{035}$ | $c_{135}$ | $c_{235}$ | $c_{335}$ | $c_{345}$ | $c_{355}$ | $c_{356}$ |
| $vS_{36}$ | $c_{036}$ | $c_{136}$ | $c_{236}$ | $c_{336}$ | $c_{346}$ | $c_{356}$ | $c_{366}$ |
| $vS_{44}$ | $c_{044}$ | $c_{144}$ | $c_{244}$ | $c_{344}$ | $c_{444}$ | $c_{445}$ | $c_{446}$ |
| $vS_{45}$ | $c_{045}$ | $c_{145}$ | $c_{245}$ | $c_{345}$ | $c_{445}$ | $c_{455}$ | $c_{456}$ |
| $vS_{46}$ | $c_{046}$ | $c_{146}$ | $c_{246}$ | $c_{346}$ | $c_{446}$ | $c_{456}$ | $c_{466}$ |
| $vS_{55}$ | $c_{055}$ | $c_{155}$ | $c_{255}$ | $c_{355}$ | $c_{455}$ | $c_{555}$ | $c_{556}$ |
| $vS_{56}$ | $c_{056}$ | $c_{156}$ | $c_{256}$ | $c_{356}$ | $c_{456}$ | $c_{556}$ | $c_{566}$ |
| $vS_{66}$ | $c_{066}$ | $c_{166}$ | $c_{266}$ | $c_{366}$ | $c_{466}$ | $c_{566}$ | $c_{666}$ |

Table 2. Third order derivatives of the free energy for a thermodynamic solid. $T$ and $\sigma_i$ are the independent variables. Exchanging the order of differentiation does not change the entry so for example $c_{033}$ is the entry in the 5th column and 5th row and also the 2nd column and 20th row. The double 33's implies it appears twice. This table may also be used as a continuation of Table 1 for all Jacobian calculations.

Table 3.  Orthorhombic Crystal Symmetry: second and third order derivatives of the free energy for a crystal restricted to orthorhombic symmetry with $T$ and $\sigma_i$ as the independent variables.

| f | $\partial/\partial T$ | $\partial/\partial \sigma_1\vert_{T,\sigma_1'}$ | $\partial/\partial \sigma_2\vert_{T,\sigma_2'}$ | $\partial/\partial \sigma_3\vert_{T,\sigma_3'}$ | $\partial/\partial \sigma_4\vert_{T,\sigma_4'}$ | $\partial/\partial \sigma_5\vert_{T,\sigma_5'}$ | $\partial/\partial \sigma_6\vert_{T,\sigma_6'}$ |
|---|---|---|---|---|---|---|---|
| g | -s | $-\Omega_1$ | $-\Omega_2$ | $-\Omega_3$ | $-\Omega_4$ | $-\Omega_5$ | $-\Omega_6$ |
| s | $C_\sigma/T$ | $v\beta_1$ | $v\beta_2$ | $v\beta_3$ | 0 | 0 | 0 |
| $\Omega_1$ | $v\beta_1$ | $vS_{11}$ | $vS_{12}$ | $vS_{13}$ | 0 | 0 | 0 |
| $\Omega_2$ | $v\beta_2$ | $vS_{12}$ | $vS_{22}$ | $vS_{23}$ | 0 | 0 | 0 |
| $\Omega_3$ | $v\beta_3$ | $vS_{13}$ | $vS_{23}$ | $vS_{33}$ | 0 | 0 | 0 |
| $\Omega_4$ | 0 | 0 | 0 | 0 | $vS_{44}$ | 0 | 0 |
| $\Omega_5$ | 0 | 0 | 0 | 0 | 0 | $vS_{55}$ | 0 |
| $\Omega_6$ | 0 | 0 | 0 | 0 | 0 | 0 | $vS_{66}$ |
| $C_\sigma/T$ | $c_{000}$ | $c_{001}$ | $c_{002}$ | $c_{003}$ | 0 | 0 | 0 |
| $v\beta_1$ | $c_{001}$ | $c_{011}$ | $c_{012}$ | $c_{013}$ | 0 | 0 | 0 |
| $v\beta_2$ | $c_{002}$ | $c_{012}$ | $c_{022}$ | $c_{023}$ | 0 | 0 | 0 |
| $v\beta_3$ | $c_{003}$ | $c_{013}$ | $c_{023}$ | $c_{033}$ | 0 | 0 | 0 |
| $vS_{11}$ | $c_{011}$ | $c_{111}$ | $c_{112}$ | $c_{113}$ | 0 | 0 | 0 |
| $vS_{12}$ | $c_{012}$ | $c_{112}$ | $c_{122}$ | $c_{123}$ | 0 | 0 | 0 |
| $vS_{13}$ | $c_{013}$ | $c_{113}$ | $c_{123}$ | $c_{133}$ | 0 | 0 | 0 |
| $vS_{22}$ | $c_{022}$ | $c_{122}$ | $c_{222}$ | $c_{223}$ | 0 | 0 | 0 |
| $vS_{23}$ | $c_{023}$ | $c_{123}$ | $c_{223}$ | $c_{233}$ | 0 | 0 | 0 |
| $vS_{33}$ | $c_{033}$ | $c_{133}$ | $c_{233}$ | $c_{333}$ | 0 | 0 | 0 |
| $vS_{44}$ | 0 | 0 | 0 | 0 | $c_{444}$ | 0 | 0 |
| $vS_{55}$ | 0 | 0 | 0 | 0 | 0 | $c_{555}$ | 0 |
| $vS_{66}$ | 0 | 0 | 0 | 0 | 0 | 0 | $c_{666}$ |

Table 3.  The orthorhombic crystal has 9 independent elastic compliances and 3 linear thermal expansion coefficients.  All the shear thermal expansion coefficients are zero.  The zero's in the upper part of Table 3 appear in the third order derivatives because of double or triple entries in Table 2.

Table 4.  Tetragonal Crystal Symmetry: Second and third order derivatives of the free energy for a crystal restricted to tetragonal crystal symmetry with $T$ and $\sigma_i$ as the independent variables.

| f | $\partial/\partial T\|_\sigma$ | $\partial/\partial \sigma_1\|_{T,\sigma_1'}$ | $\partial/\partial \sigma_2\|_{T,\sigma_2'}$ | $\partial/\partial \sigma_3\|_{T,\sigma_3'}$ | $\partial/\partial \sigma_4\|_{T,\sigma_4'}$ | $\partial/\partial \sigma_5\|_{T,\sigma_5'}$ | $\partial/\partial \sigma_6\|_{T,\sigma_6'}$ |
|---|---|---|---|---|---|---|---|
| g | -s | $-\Omega_1$ | $-\Omega_2$ | $-\Omega_3$ | $-\Omega_4$ | $-\Omega_5$ | $-\Omega_6$ |
| s | $C_\sigma/T$ | $v\beta_1$ | $v\beta_1$ | $v\beta_3$ | 0 | 0 | 0 |
| $\Omega_1$ | $v\beta_1$ | $vS_{11}$ | $vS_{12}$ | $vS_{13}$ | 0 | 0 | 0 |
| $\Omega_2$ | $v\beta_1$ | $vS_{12}$ | $vS_{11}$ | $vS_{13}$ | 0 | 0 | 0 |
| $\Omega_3$ | $v\beta_3$ | $vS_{13}$ | $vS_{13}$ | $vS_{33}$ | 0 | 0 | 0 |
| $\Omega_4$ | 0 | 0 | 0 | 0 | $vS_{44}$ | 0 | 0 |
| $\Omega_5$ | 0 | 0 | 0 | 0 | 0 | $vS_{44}$ | 0 |
| $\Omega_6$ | 0 | 0 | 0 | 0 | 0 | 0 | $vS_{66}$ |
| $C_\sigma/T$ | $c_{000}$ | $c_{001}$ | $c_{002}$ | $c_{003}$ | 0 | 0 | 0 |
| $v\beta_1$ | $c_{001}$ | $c_{011}$ | $c_{012}$ | $c_{013}$ | 0 | 0 | 0 |
| $v\beta_3$ | $c_{003}$ | $c_{013}$ | $c_{023}$ | $c_{033}$ | 0 | 0 | 0 |
| $vS_{11}$ | $c_{011}$ | $c_{111}$ | $c_{112}$ | $c_{113}$ | 0 | 0 | 0 |
| $vS_{12}$ | $c_{012}$ | $c_{112}$ | $c_{122}$ | $c_{123}$ | 0 | 0 | 0 |
| $vS_{13}$ | $c_{013}$ | $c_{113}$ | $c_{123}$ | $c_{133}$ | 0 | 0 | 0 |
| $vS_{33}$ | $c_{033}$ | $c_{133}$ | $c_{233}$ | $c_{333}$ | 0 | 0 | 0 |
| $vS_{44}$ | 0 | 0 | 0 | 0 | 0 | 0 | 0 |
| $vS_{66}$ | 0 | 0 | 0 | 0 | 0 | 0 | $c_{666}$ |

Table 4.  Tetragonal symmetry for classes 4*mm*, $\bar{4}2m$, 422, 4/*mmm* for this crystal class there are 6 independent elastic compliances and 2 thermal expansion coefficients.  Again, many of the zero entries must appear twice or three times in the Table 4.

Table 5. Cubic Crystal Symmetry: Second and third order derivatives of the free energy for a crystal restricted to cubic crystal symmetry with $T$ and $\sigma_i$ as the independent variables.

| f | $\partial/\partial T\vert_\sigma$ | $\partial/\partial\sigma_1\vert_{T,\sigma_1'}$ | $\partial/\partial\sigma_2\vert_{T,\sigma_2'}$ | $\partial/\partial\sigma_3\vert_{T,\sigma_3'}$ | $\partial/\partial\sigma_4\vert_{T,\sigma_4'}$ | $\partial/\partial\sigma_5\vert_{T,\sigma_5'}$ | $\partial/\partial\sigma_6$ |
|---|---|---|---|---|---|---|---|
| g | $-s$ | $-\Omega_1$ | $-\Omega_2$ | $-\Omega_3$ | $-\Omega_4$ | $-\Omega_5$ | $-\Omega_6$ |
| s | $c_\sigma/T$ | $v\beta_1$ | $v\beta_1$ | $v\beta_1$ | 0 | 0 | 0 |
| $\Omega_1$ | $v\beta_1$ | $vS_{11}$ | $vS_{12}$ | $vS_{12}$ | 0 | 0 | 0 |
| $\Omega_2$ | $v\beta_1$ | $vS_{12}$ | $vS_{11}$ | $vS_{12}$ | 0 | 0 | 0 |
| $\Omega_3$ | $v\beta_1$ | $vS_{12}$ | $vS_{12}$ | $vS_{11}$ | 0 | 0 | 0 |
| $\Omega_4$ | 0 | 0 | 0 | 0 | $vS_{44}$ | 0 | 0 |
| $\Omega_5$ | 0 | 0 | 0 | 0 | 0 | $vS_{44}$ | 0 |
| $\Omega_6$ | 0 | 0 | 0 | 0 | 0 | 0 | $vS_{44}$ |
| $c_\sigma/T$ | $c_{000}$ | $c_{001}$ | $c_{001}$ | $c_{001}$ | 0 | 0 | 0 |
| $v\beta_1$ | $c_{001}$ | $c_{011}$ | $c_{011}$ | $c_{011}$ | 0 | 0 | 0 |
| $vS_{11}$ | $c_{011}$ | $c_{111}$ | $c_{111}$ | $c_{111}$ | 0 | 0 | 0 |
| $vS_{12}$ | $c_{011}$ | $c_{111}$ | $c_{111}$ | $c_{111}$ | 0 | 0 | 0 |
| $vS_{44}$ | 0 | 0 | 0 | 0 | 0 | 0 | 0 |

Table 5. Cubic symmetry applied to a crystal has 3 independent elastic compliances and a single thermal expansion coefficient. Again, many of the zero entries would have appeared twice or three times from Table 2. Note that $vS_{44}$ has no temperature dependence and its derivative with respect to the variable T is zero because there is no thermal expansion of $\beta_4$ at zero shear stress. The shear thermal expansion coefficient for a cell that remains cubic is zero. Also, the T dependence of $S_{11}$ is the same as $S_{12}$. Also, the $\sigma_1$ dependence of $S_{11}$ is the same as $S_{12}$. Measurements are generally not of the isothermal shear compliances but rather the adiabatic compliance. See the text. The isothermal and adiabatic shear compliances are the same at zero stress. Adiabatic measurements are made from fast wave propagation which does not allow time for thermal equilibrium.

*Acknowledgements*

I'd like to thank: J. C. Lambropoulos, S. P. Burns, J. C. M. Li, E. Burnham-Fay, D. N. Polsin, S. X. Hu, G. W. Collins, J. R. Rygg and T. R. Boehly for discussions and the Department of Energy through Basic Energy Sciences, Materials Science who supported earlier work related to this topic.

*Appendix A*

Consider the following 8 equations:

$$y_1 = y_1(x_1, x_2, x_3, x_4, x_5, x_6, x_7)$$
$$y_2 = y_2(x_1, x_2, x_3, x_4, x_5, x_6, x_7)$$
$$y_3 = y_3(x_1, x_2, x_3, x_4, x_5, x_6, x_7)$$
$$y_4 = y_4(x_1, x_2, x_3, x_4, x_5, x_6, x_7)$$
$$y_5 = y_5(x_1, x_2, x_3, x_4, x_5, x_6, x_7)$$
$$y_6 = y_6(x_1, x_2, x_3, x_4, x_5, x_6, x_7)$$
$$y_7 = y_7(x_1, x_2, x_3, x_4, x_5, x_6, x_7)$$
$$y_8 = y_8(x_1, x_2, x_3, x_4, x_5, x_6, x_7)$$

Their chain rule derivatives are

$$dy_1 = \frac{\partial y_1}{\partial x_1} dx_1 + \frac{\partial y_1}{\partial x_2} dx_2 + \ldots \frac{\partial y_1}{\partial x_7} dx_7$$

$$dy_2 = \frac{\partial y_2}{\partial x_1} dx_1 + \frac{\partial y_2}{\partial x_2} dx_2 + \ldots \frac{\partial y_2}{\partial x_7} dx_7$$

$$\vdots$$

$$dy_8 = \frac{\partial y_8}{\partial x_1} dx_1 + \frac{\partial y_8}{\partial x_2} dx_2 + \ldots \frac{\partial y_8}{\partial x_7} dx_7$$

Let $y_1$ and $y_2$ vary while holding $y_3, y_4, \ldots, y_8$ constant. The following 8 homogeneous equations are found:

$$0 = -dy_1 + \frac{\partial y_1}{\partial x_1} dx_1 + \frac{\partial y_1}{\partial x_2} dx_2 + \ldots \frac{\partial y_1}{\partial x_7} dx_7$$

$$0 = -dy_2 + \frac{\partial y_2}{\partial x_1} dx_1 + \frac{\partial y_2}{\partial x_2} dx_2 + \ldots \frac{\partial y_2}{\partial x_7} dx_7$$

$$0 = 0 + \frac{\partial y_3}{\partial x_1} dx_1 + \frac{\partial y_3}{\partial x_2} dx_2 + \ldots \frac{\partial y_3}{\partial x_7} dx_7$$

$$\vdots$$

$$0 = 0 + \frac{\partial y_8}{\partial x_1} dx_1 + \frac{\partial y_8}{\partial x_2} dx_2 + \ldots \frac{\partial y_8}{\partial x_7} dx_7$$

or

$$\begin{pmatrix} 0 \\ 0 \\ 0 \\ 0 \\ 0 \\ 0 \\ 0 \\ 0 \end{pmatrix} = \begin{pmatrix} -dy_1 & \frac{\partial y_1}{\partial x_1} & \frac{\partial y_1}{\partial x_2} & \cdots & \frac{\partial y_1}{\partial x_7} \\ -dy_2 & \frac{\partial y_2}{\partial x_1} & \frac{\partial y_2}{\partial x_2} & \cdots & \frac{\partial y_2}{\partial x_7} \\ 0 & \frac{\partial y_3}{\partial x_1} & \frac{\partial y_3}{\partial x_2} & \cdots & \frac{\partial y_3}{\partial x_7} \\ 0 & \cdots & & & \\ 0 & \cdots & & & \\ 0 & \cdots & & & \\ 0 & \cdots & & & \\ 0 & \frac{\partial y_8}{\partial x_1} & \frac{\partial y_8}{\partial x_2} & \cdots & \frac{\partial y_8}{\partial x_7} \end{pmatrix} \bullet \begin{pmatrix} 1 \\ dx_1 \\ dx_2 \\ dx_3 \\ dx_4 \\ dx_5 \\ dx_6 \\ dx_6 \\ dx_7 \end{pmatrix}$$

The determinate of the matrix must be zero for a solution

$$0 = \det \begin{vmatrix} -dy_1 & \frac{\partial y_1}{\partial x_1} & \frac{\partial y_1}{\partial x_2} & \cdots & \frac{\partial y_1}{\partial x_7} \\ -dy_2 & \frac{\partial y_2}{\partial x_1} & \frac{\partial y_2}{\partial x_2} & \cdots & \frac{\partial y_2}{\partial x_7} \\ 0 & \frac{\partial y_3}{\partial x_1} & \frac{\partial y_3}{\partial x_2} & \cdots & \frac{\partial y_3}{\partial x_7} \\ 0 & \cdots & & & \\ 0 & \cdots & & & \\ 0 & \cdots & & & \\ 0 & \cdots & & & \\ 0 & \frac{\partial y_8}{\partial x_1} & \frac{\partial y_8}{\partial x_2} & \cdots & \frac{\partial y_8}{\partial x_7} \end{vmatrix}$$

Evaluating the determinate using Cramer's Rule gives

$$0 = -dy_1 \left( \det \begin{vmatrix} \frac{\partial y_2}{\partial x_1} & \frac{\partial y_2}{\partial x_2} & \cdots & \frac{\partial y_2}{\partial x_7} \\ \frac{\partial y_3}{\partial x_1} & \frac{\partial y_3}{\partial x_2} & \cdots & \frac{\partial y_3}{\partial x_7} \\ & & & \\ \frac{\partial y_8}{\partial x_1} & \frac{\partial y_8}{\partial x_2} & \cdots & \frac{\partial y_8}{\partial x_7} \end{vmatrix} \right) + dy_2 \left( \det \begin{vmatrix} \frac{\partial y_1}{\partial x_1} & \frac{\partial y_1}{\partial x_2} & \cdots & \frac{\partial y_1}{\partial x_7} \\ \frac{\partial y_3}{\partial x_1} & \frac{\partial y_3}{\partial x_2} & \cdots & \frac{\partial y_3}{\partial x_7} \\ & & & \\ \frac{\partial y_8}{\partial x_1} & \frac{\partial y_8}{\partial x_2} & \cdots & \frac{\partial y_8}{\partial x_7} \end{vmatrix} \right) \text{ so}$$

$$\left.\frac{\partial y_1}{\partial y_2}\right|_{y_3,y_4,\ldots y_8} = \frac{\det\begin{vmatrix} \frac{\partial y_1}{\partial x_1} & \frac{\partial y_1}{\partial x_2} & \cdots & \frac{\partial y_1}{\partial x_7} \\ \frac{\partial y_3}{\partial x_1} & \frac{\partial y_3}{\partial x_2} & \cdots & \frac{\partial y_3}{\partial x_7} \\ & & & \\ \frac{\partial y_8}{\partial x_1} & \frac{\partial y_8}{\partial x_2} & \cdots & \frac{\partial y_8}{\partial x_7} \end{vmatrix}}{\det\begin{vmatrix} \frac{\partial y_2}{\partial x_1} & \frac{\partial y_2}{\partial x_2} & \cdots & \frac{\partial y_2}{\partial x_7} \\ \frac{\partial y_3}{\partial x_1} & \frac{\partial y_3}{\partial x_2} & \cdots & \frac{\partial y_3}{\partial x_7} \\ & & & \\ \frac{\partial y_8}{\partial x_1} & \frac{\partial y_8}{\partial x_2} & \cdots & \frac{\partial y_8}{\partial x_7} \end{vmatrix}}$$

or

$$\left.\frac{\partial y_1}{\partial y_2}\right|_{y_3,y_4,\ldots y_8} = \frac{J(y_1,y_3,y_4,y_5,y_6,y_7,y_8)}{J(y_2,y_3,y_4,y_5,y_6,y_7,y_8)}$$

Tables 1-5 in the text are all written to be in Jacobian format so any partial derivative can be found. For example

$$\left.\frac{\partial y_1}{\partial y_2}\right|_{y_3,y_4,\ldots y_8}$$

is formed by reading the rows in the tables. MatLab works well in evaluating the 7X7 determinates. For example:

```
%% Evaluation of determinates from Jacobians in Strain Volume Thermodynamics
%% c is constant stress heat capacity; T is absolute temperature;
%% betai are crystallographic thermal expansion coefficients; v is volume
%% per unit mass; sij are elastic compliances.

syms c T v beta1 beta2 beta3 beta4 beta5 beta6 s11 s12 s13 s14 s15 s16
M1=[c/T, v*beta1, v*beta2, v*beta3, v*beta4, v*beta5, v*beta6;
    1, 0, 0, 0, 0, 0, 0;
    0, 0, 1, 0, 0, 0, 0;
    0, 0, 0, 1, 0, 0, 0;
    0, 0, 0, 0, 1, 0, 0;
    0, 0, 0, 0, 0, 1, 0;
    0, 0, 0, 0, 0, 0, 1]
det(M1)
```

```
M2= [ v*beta1, v*s11, v*s12, v*s13, v*s14, v*s15, v*s16;
    1, 0, 0, 0, 0, 0, 0;
    0, 0, 1, 0, 0, 0, 0;
    0, 0, 0, 1, 0, 0, 0;
    0, 0, 0, 0, 1, 0, 0;
    0, 0, 0, 0, 0, 1, 0;
    0, 0, 0, 0, 0, 0, 1]

det(M2)

det(M1)/det(M2)
```

Which has a MatLab evaluation as:
M1 =
[ c/T, beta1*v, beta2*v, beta3*v, beta4*v, beta5*v, beta6*v]
[ 1,   0,   0,   0,   0,   0,   0]
[ 0,   0,   1,   0,   0,   0,   0]
[ 0,   0,   0,   1,   0,   0,   0]
[ 0,   0,   0,   0,   1,   0,   0]
[ 0,   0,   0,   0,   0,   1,   0]
[ 0,   0,   0,   0,   0,   0,   1]
ans =
-beta1*v

M2 =
[ beta1*v, s11*v, s12*v, s13*v, s14*v, s15*v, s16*v]
[ 1,   0,   0,   0,   0,   0,   0]
[ 0,   0,   1,   0,   0,   0,   0]
[ 0,   0,   0,   1,   0,   0,   0]
[ 0,   0,   0,   0,   1,   0,   0]
[ 0,   0,   0,   0,   0,   1,   0]
[ 0,   0,   0,   0,   0,   0,   1]
ans =
-s11*v

ans =
beta1/s11

so the above expression used in equation (17) is evaluated as

$$\left.\frac{\partial s}{\partial \Omega_1}\right|_{T,\sigma_2,\ldots\sigma_6} = \frac{J(s,T,\sigma_2,\sigma_3,\sigma_4,\sigma_5,\sigma_6)}{J(\Omega_1,T,\sigma_2,\sigma_3,\sigma_4,\sigma_5,\sigma_6)} = \frac{\beta_1}{S_{11}}$$

**Figures:**

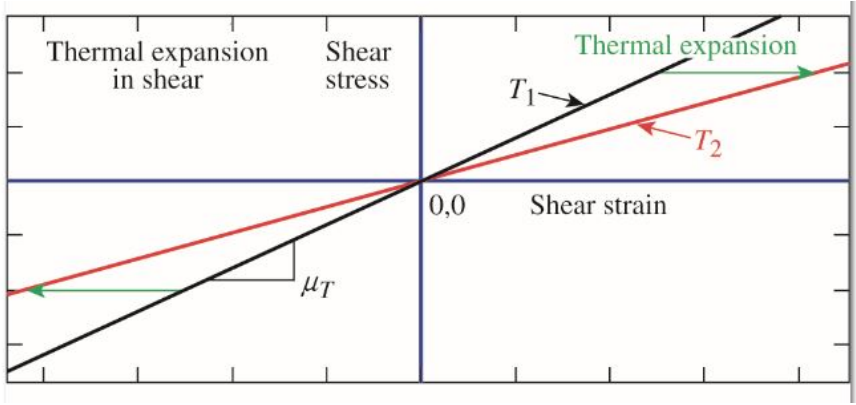

Figure 1. A schematic of the elastic shear stress, $\sigma_4$, versus the elastic shear strain, $\varepsilon_4$, showing two isotherms. The isothermal lines pass through the point of zero shear stress, zero shear strain. If $T_2$ is larger than $T_1$ then the first quadrant has positive shear thermal expansion coefficients and the third quadrant has negative values. The origin has zero thermal shear thermal expansion coefficients.

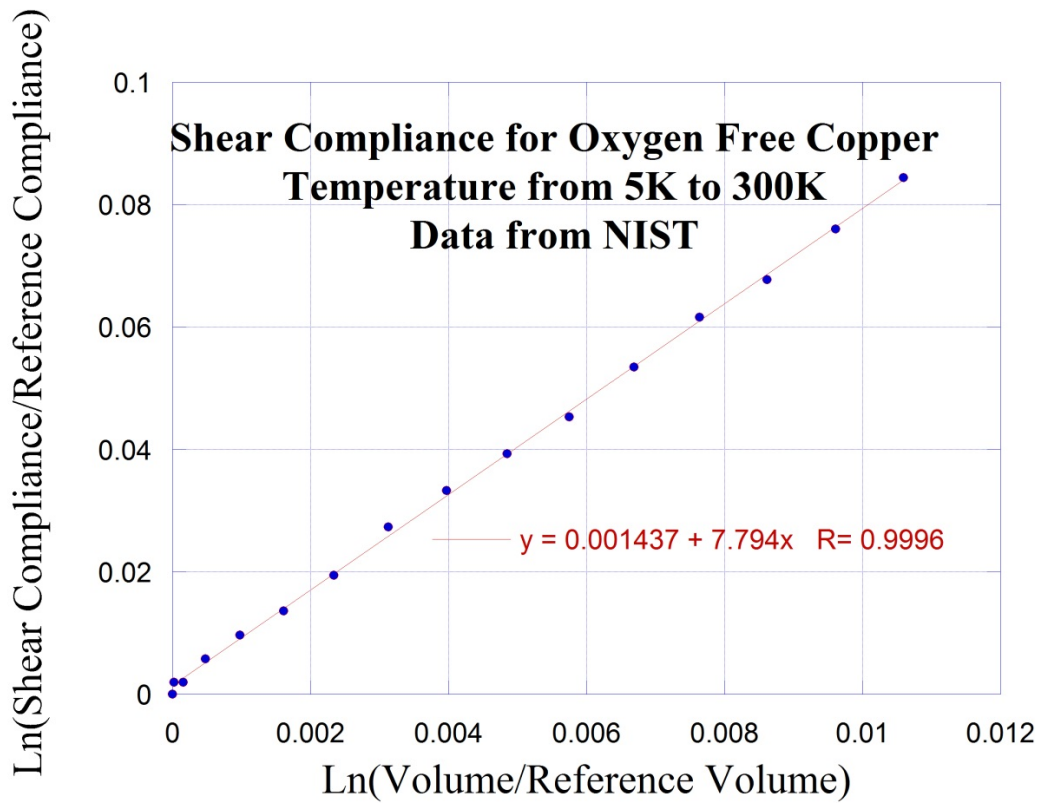

Figure 2. A plot of $\ell n$(Compliance) versus $\ell n$(Volume) from measured shear moduli over a reference modulus versus the measured volume over a reference volume is displayed. $T$ is the parameter and goes from 5K to 300K. See equation (20). The data has been curve fit with a linear straight line that is not forced to go through 0, 0. Data is from reference [38].

Figure Source:

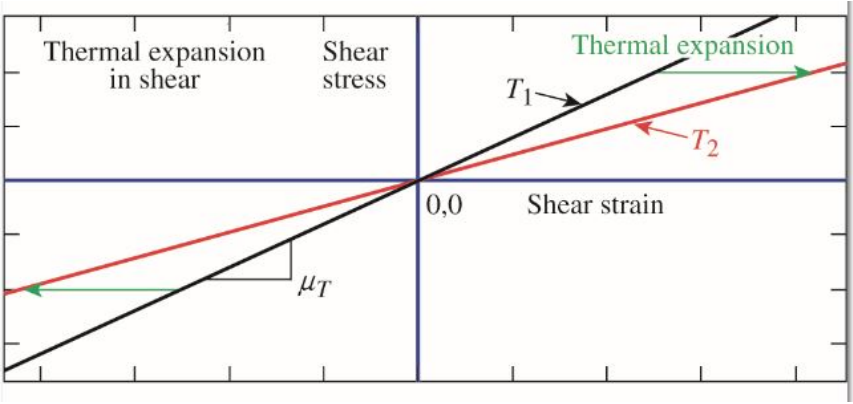

Fig 1

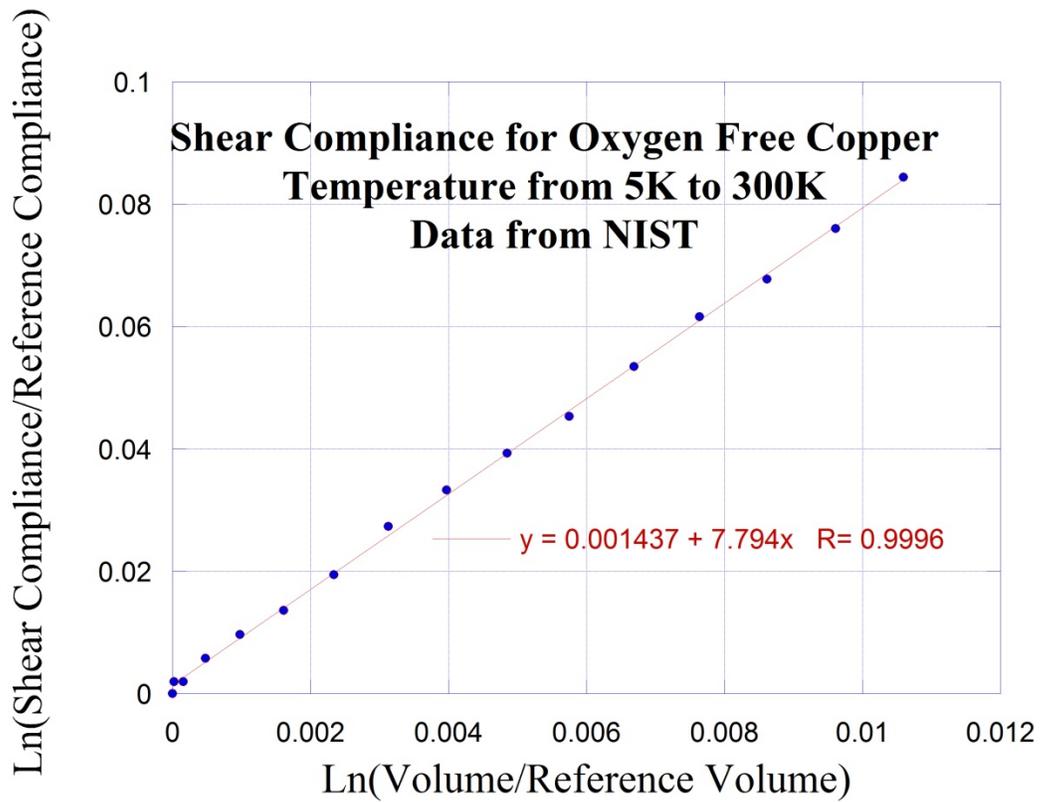

Fig 2